\documentstyle[12pt,psfig]{article}

\catcode`\@=11
\long\def\@makefntext#1{ %\parindent 1em
\protect\noindent \hbox to 3.2pt {\hskip-.9pt  
$^{{\ninerm\@thefnmark}}$\hfil}#1\hfill} %can be used 

\def\thefootnote{\fnsymbol{footnote}}
 \def\@makefnmark{\hbox to 0pt{$^{\@thefnmark}$\hss}}  %original 
	
\def\ps@myheadings{\let\@mkboth\@gobbletwo
\def\@oddhead{\hbox{} %\sl
\rightmark\hfil\ninerm\thepage}   
\def\@oddfoot{}\def\@evenhead{\ninerm\thepage\hfil %\sl
\leftmark\hbox{}}\def\@evenfoot{}
\def\sectionmark##1{}\def\subsectionmark##1{}}

\begin{document}

\newcommand{\gsim}{\mbox{ \raisebox{-1.0ex}{$\stackrel{\textstyle >}
{\textstyle \sim}$ }}}
\newcommand{\lsim}{\mbox{ \raisebox{-1.0ex}{$\stackrel{\textstyle <}
{\textstyle \sim}$ }}}

%----------------------------PROCSLA.STY---------------------------------------
\newcommand{\symbolfootnote}{\renewcommand{\thefootnote}
	{\fnsymbol{footnote}}}
\renewcommand{\thefootnote}{\fnsymbol{footnote}}
\newcommand{\alphfootnote}
	{\setcounter{footnote}{0}
	 \renewcommand{\thefootnote}{\sevenrm\alph{footnote}}}

%------------------------------------------------------------------------------
%NEW DEFINED SECTION COMMANDS 
\newcounter{sectionc}\newcounter{subsectionc}\newcounter{subsubsectionc}
\renewcommand{\section}[1] {\vspace{0.6cm}\addtocounter{sectionc}{1} 
\setcounter{subsectionc}{0}\setcounter{subsubsectionc}{0}\noindent 
	{\bf\thesectionc. #1}\par\vspace{0.4cm}}
\renewcommand{\subsection}[1] {\vspace{0.6cm}\addtocounter{subsectionc}{1} 
	\setcounter{subsubsectionc}{0}\noindent 
	{\it\thesectionc.\thesubsectionc. #1}\par\vspace{0.4cm}}
\renewcommand{\subsubsection}[1] {\vspace{0.6cm}\addtocounter{subsubsectionc}{1}
	\noindent {\rm\thesectionc.\thesubsectionc.\thesubsubsectionc. 
	#1}\par\vspace{0.4cm}}
\newcommand{\nonumsection}[1] {\vspace{0.6cm}\noindent{\bf #1}
	\par\vspace{0.4cm}}
					         
%NEW MACRO TO HANDLE APPENDICES
\newcounter{appendixc}
\newcounter{subappendixc}[appendixc]
\newcounter{subsubappendixc}[subappendixc]
\renewcommand{\thesubappendixc}{\Alph{appendixc}.\arabic{subappendixc}}
\renewcommand{\thesubsubappendixc}
	{\Alph{appendixc}.\arabic{subappendixc}.\arabic{subsubappendixc}}

\renewcommand{\appendix}[1] {\vspace{0.6cm}
        \refstepcounter{appendixc}
        \setcounter{figure}{0}
        \setcounter{table}{0}
        \setcounter{equation}{0}
        \renewcommand{\thefigure}{\Alph{appendixc}.\arabic{figure}}
        \renewcommand{\thetable}{\Alph{appendixc}.\arabic{table}}
        \renewcommand{\theappendixc}{\Alph{appendixc}}
        \renewcommand{\theequation}{\Alph{appendixc}.\arabic{equation}}
%       \noindent{\bf Appendix \theappendixc. #1}\par\vspace{0.4cm}}
        \noindent{\bf Appendix \theappendixc #1}\par\vspace{0.4cm}}
\newcommand{\subappendix}[1] {\vspace{0.6cm}
        \refstepcounter{subappendixc}
        \noindent{\bf Appendix \thesubappendixc. #1}\par\vspace{0.4cm}}
\newcommand{\subsubappendix}[1] {\vspace{0.6cm}
        \refstepcounter{subsubappendixc}
        \noindent{\it Appendix \thesubsubappendixc. #1}
	\par\vspace{0.4cm}}

%------------------------------------------------------------------------------
%MARCO FOR ABSTRACT BLOCK
\def\abstracts#1{{
	\centering{\begin{minipage}{30pc}\tenrm\baselineskip=12pt\noindent
	\centerline{\tenrm ABSTRACT}\vspace{0.3cm}
	\parindent=0pt #1
	\end{minipage} }\par}} 

%------------------------------------------------------------------------------
%NEW MACRO FOR BIBLIOGRAPHY
\newcommand{\bibit}{\it}
\newcommand{\bibbf}{\bf}
\renewenvironment{thebibliography}[1]
	{\begin{list}{\arabic{enumi}.}
	{\usecounter{enumi}\setlength{\parsep}{0pt}
%1.25cm IS STRICTLY FOR PROCSLA.TEX ONLY
\setlength{\leftmargin 1.25cm}{\rightmargin 0pt}
%0.52cm IS FOR NEW DATA FILES
%\setlength{\leftmargin 0.52cm}{\rightmargin 0pt}
	 \setlength{\itemsep}{0pt} \settowidth
	{\labelwidth}{#1.}\sloppy}}{\end{list}}

%------------------------------------------------------------------------------
%FOLLOWING THREE COMMANDS ARE FOR 'LIST' COMMAND.
\topsep=0in\parsep=0in\itemsep=0in
\parindent=1.5pc

%LIST ENVIRONMENTS
\newcounter{itemlistc}
\newcounter{romanlistc}
\newcounter{alphlistc}
\newcounter{arabiclistc}
\newenvironment{itemlist}
    	{\setcounter{itemlistc}{0}
	 \begin{list}{$\bullet$}
	{\usecounter{itemlistc}
	 \setlength{\parsep}{0pt}
	 \setlength{\itemsep}{0pt}}}{\end{list}}

\newenvironment{romanlist}
	{\setcounter{romanlistc}{0}
	 \begin{list}{$($\roman{romanlistc}$)$}
	{\usecounter{romanlistc}
	 \setlength{\parsep}{0pt}
	 \setlength{\itemsep}{0pt}}}{\end{list}}

\newenvironment{alphlist}
	{\setcounter{alphlistc}{0}
	 \begin{list}{$($\alph{alphlistc}$)$}
	{\usecounter{alphlistc}
	 \setlength{\parsep}{0pt}
	 \setlength{\itemsep}{0pt}}}{\end{list}}

\newenvironment{arabiclist}
	{\setcounter{arabiclistc}{0}
	 \begin{list}{\arabic{arabiclistc}}
	{\usecounter{arabiclistc}
	 \setlength{\parsep}{0pt}
	 \setlength{\itemsep}{0pt}}}{\end{list}}

%------------------------------------------------------------------------------
%FIGURE CAPTION
\newcommand{\fcaption}[1]{
        \refstepcounter{figure}
        \setbox\@tempboxa = \hbox{\tenrm Fig.~\thefigure. #1}
        \ifdim \wd\@tempboxa > 6in
           {\begin{center}
        \parbox{6in}{\tenrm\baselineskip=12pt Fig.~\thefigure. #1 }
            \end{center}}
        \else
             {\begin{center}
             {\tenrm Fig.~\thefigure. #1}
              \end{center}}
        \fi}

%TABLE CAPTION
\newcommand{\tcaption}[1]{
        \refstepcounter{table}
        \setbox\@tempboxa = \hbox{\tenrm Table~\thetable. #1}
        \ifdim \wd\@tempboxa > 6in
           {\begin{center}
        \parbox{6in}{\tenrm\baselineskip=12pt Table~\thetable. #1 }
            \end{center}}
        \else
             {\begin{center}
             {\tenrm Table~\thetable. #1}
              \end{center}}
        \fi}

%------------------------------------------------------------------------------
%ACKNOWLEDGEMENT: this portion is from John Hershberger
\def\@citex[#1]#2{\if@filesw\immediate\write\@auxout
	{\string\citation{#2}}\fi
\def\@citea{}\@cite{\@for\@citeb:=#2\do
	{\@citea\def\@citea{,}\@ifundefined
	{b@\@citeb}{{\bf ?}\@warning
	{Citation `\@citeb' on page \thepage \space undefined}}
	{\csname b@\@citeb\endcsname}}}{#1}}

\newif\if@cghi
\def\cite{\@cghitrue\@ifnextchar [{\@tempswatrue
	\@citex}{\@tempswafalse\@citex[]}}
\def\citelow{\@cghifalse\@ifnextchar [{\@tempswatrue
	\@citex}{\@tempswafalse\@citex[]}}
\def\@cite#1#2{{$\null^{#1}$\if@tempswa\typeout
	{IJCGA warning: optional citation argument 
	ignored: `#2'} \fi}}
\newcommand{\citeup}{\cite}

%------------------------------------------------------------------------------
%FOR FNSYMBOL FOOTNOTE AND ALPH{FOOTNOTE} 
\def\fnm#1{$^{\mbox{\scriptsize #1}}$}
\def\fnt#1#2{\footnotetext{\kern-.3em
	{$^{\mbox{\sevenrm #1}}$}{#2}}}

%------------------------------------------------------------------------------
\font\twelvebf=cmbx10 scaled\magstep 1
\font\twelverm=cmr10 scaled\magstep 1
\font\twelveit=cmti10 scaled\magstep 1
\font\elevenbfit=cmbxti10 scaled\magstephalf
\font\elevenbf=cmbx10 scaled\magstephalf
\font\elevenrm=cmr10 scaled\magstephalf
\font\elevenit=cmti10 scaled\magstephalf
\font\bfit=cmbxti10
\font\tenbf=cmbx10
\font\tenrm=cmr10
\font\tenit=cmti10
\font\ninebf=cmbx9
\font\ninerm=cmr9
\font\nineit=cmti9
\font\eightbf=cmbx8
\font\eightrm=cmr8
\font\eightit=cmti8

%----------------------START OF DATA FILE------------------------------
\begin{flushright}
  \begin{tabular}[t]{l} 
  KEK-TH-508\\
  KEK Preprint 96-159\\
  January 1997
 \end{tabular}
 \end{flushright}
\vspace*{0.5cm}
%\centerline{\tenbf WORLD SCIENTIFIC PUBLISHING COMPANY}
%\baselineskip=22pt
\centerline{\tenbf PHENOMENOLOGY OF THE HIGGS SECTOR IN}
\baselineskip=16pt
\centerline{\tenbf SUPERSYMMETRIC STANDARD MODEL
\footnote{
    Talk presented
    at International Workshop on Frontiers in
    Quantum Field Theory, August 11-18, 1996, Urumqi, China.
}}
%\centerline{\ninerm (For 20\% Reduction to 6 in. $\times$ 8.5 in. Trim Size)}
\vspace{0.8cm}
\centerline{\tenrm YASUHIRO OKADA}
\baselineskip=13pt
\centerline{\tenit National Laboratory for High Energy Physics (KEK)}
\baselineskip=12pt
\centerline{\tenit Oho 1-1, Tsukuba 305, Japan}
\vspace{0.3cm}
%\centerline{\tenrm and}
%\vspace{0.3cm}
%\centerline{\tenrm SECOND AUTHOR'S NAME}
%\baselineskip=13pt
%\centerline{\tenit Group, Company, Address, City, State ZIP/Zone, Country}
\vspace{0.9cm}
\abstracts{Several topics related to phenomenology of 
the Higgs sector in the supersymmetric standard model  
are reviewed.
The upper bound of the lightest Higgs mass in the minimal 
supersymmetric standard model as well as extended version
of it is discussed and it is shown that an 
$e^+ e^-$ linear collider with $\sqrt{s}\sim300 - 500$ GeV
can find at least one Higgs boson in these models. 
It is also pointed out that the heavy Higgs mass scale may be
determined from measurements of the Higgs boson decay 
branching ratios even if we only discover the lightest 
Higgs boson at early stage of the linear collider 
experiment.}

\vfil
%\vspace{0.8cm}
\twelverm   %modified by CLee 23/07/93
\baselineskip=14pt

\section{Introduction}
After the discovery of top quark at Fermilab and precise measurements
of electroweak interaction at LEP and SLC experiments it has been 
more and more evident that the elementary particle physics is 
described by the Standard Model (SM). This model is based on two 
physical principles, {\it i.e.} the gauge principle and the
Higgs mechanism. Although we can understand most of the experimental
results by the $SU(3)\times SU(2)\times U(1)$ gauge symmetry,
little is known about the dynamics behind the electroweak symmetry 
breaking. Therefore exploring the Higgs sector is the most 
important issue of the current high energy physics and the primary
objective of future collider experiments.

Study of the Higgs sector is not only important to establish the 
SM but also crucial to search for physics beyond the SM. In this
respect the mass of the Higgs boson itself gives us important
information. For example, a heavy Higgs boson suggests that the
dynamics of the electroweak symmetry breaking is governed by strong
interaction. On the other hand if we assume that fundamental
interactions are described by perturbation theory up to
the Planck scale or a scale close to it, the Higgs boson is 
expected to exist below 200 GeV. Grand unified theory (GUT)
and supersymmetric (SUSY) extension of the SM
are examples of the latter case.

In this talk I would like to discuss phenomenological aspects
of the supersymmetric standard model, especially some issues
related to the Higgs sector of the SUSY model. Among various
extensions of the SM, the SUSY SM is now supposed to be the most
promising candidate of physics beyond the SM. Since early
80's SUSY extensions of the SM and GUT have been extensively
studied because SUSY is the unique symmetry to ensure the 
smallness of electroweak scale compared to the Planck scale
by cancelling the quadratic divergence of scalar mass 
renormalization. More recently, SUSY theories attracted much 
attention since three gauge coupling constants measured precisely
at LEP and SLC experiments are consistent with SUSY GUT
although the simplest non-SUSY GUT is excluded experimentally.

In the following sections, I first discuss the Higgs sector
of the minimal supersymmetric standard model (MSSM). It is
shown that the upper bound of the lightest CP-even Higgs boson
in this model is given by about 130 GeV, which is a prime
target of future experiments at LHC and $e^+e^-$ linear
colliders. Then the Higgs sector of extended version of 
the SUSY SM is reviewed. Finally, I show how the measurements
of various Higgs decay branching ratios are useful to determine
MSSM parameters in future $e^+e^-$ linear collider experiments.    
        
\section{The Higgs Sector in the MSSM}
In order to construct a SUSY version of the SM
we need to introduce a SUSY partner for each
particle of the SM. For quarks and leptons
their scalar partners, squarks and sleptons, 
are introduced. Corresponding to the  $SU(3)$,
$SU(2)$ and $U(1)$ gauge fields we need spin 
$1/2$ gauge fermions called gluino, wino
and bino respectively. 
Unlike the minimal SM the $SU(2)$-doublet 
Higgs field giving masses to up-type quarks
and that to down-type quarks and leptons
have to be introduced separately in SUSY
models, therefore the Higgs sector contains
at least two doublet Higgs fields. 
In the minimal SUSY extension, {\it i.e.}
the MSSM, we introduce two Higgs doublets and
their SUSY partners, higgsinos. The winos, bino 
and higgsinos can have mixings due to the electroweak
symmetry breaking and form four neutral Majorana 
fermions (neutralinos) and two charged Dirac
fermions (charginos). Therefore the MSSM is 
characterized as a two doublet-Higgs
SM with scalar superpartners (squarks/sleptons)
and fermionic superpartners (neutralinos/charginos).
 
Let us first discuss the MSSM Higgs sector.
The most important feature is
that the Higgs-self-coupling
constant at the tree level is
completely determined by the $SU(2)$ and $U(1)$ gauge coupling
constants. After electroweak symmetry breaking, the physical
Higgs states include two CP-even Higgs bosons ($h, H$),
one CP-odd Higgs boson ($A$) and one pair of charged Higgs
bosons ($H^\pm$) where we denote by $h$
and $H$ the lighter and heavier Higgs bosons respectively.  
Although at the tree level the
upper bound on the lightest CP-even Higgs boson mass is given
by the $Z^0$ mass, the radiative corrections
weaken this bound.\cite{OYY} The Higgs potential is given by

\begin{eqnarray}
V_{Higgs} & = & m^2_1|H_1|^2 + m_2^2|H_2|^2 - m_3^2
   (H_{1}\cdot H_{2}+\bar{H_{1}}\cdot\bar{ H_{2}})
\nonumber\\
          & & +\frac{g_2^2}{8}(\bar{H_1}\tau^aH_1
                + \bar{H_2}\tau^aH_2)^2
                + \frac{g_1^2}{8}(|H_1|^2 - |H_2|^2)^2
\nonumber\\
          & & + \Delta V,
\end{eqnarray}
where $\Delta V$ represents the contribution from one-loop diagrams.
Since the loop correction due to the top quark and its
superpartner, the stop squark, is proportional to the fourth
power of the top Yukawa coupling constant and hence is large, the
Higgs self-coupling constant is no longer determined only by
the gauge coupling constants.
The upper bound on the lightest CP-even Higgs mass ($m_h$)
can significantly increase for a reasonable choice of the
top-quark and stop-squark masses. Figure 1 shows
the upper bound on $m_h$ as a function of top-quark mass
for several choices of the stop mass and the ratio of two Higgs-boson
vacuum expectation values ($\tan\beta = \frac{<H_2^0>}{<H_1^0>})$.
%Fig1 (uppernound on the lightest Higgs mass in MSSM)
%
\begin{figure}
\begin{center}
\mbox{\psfig{figure=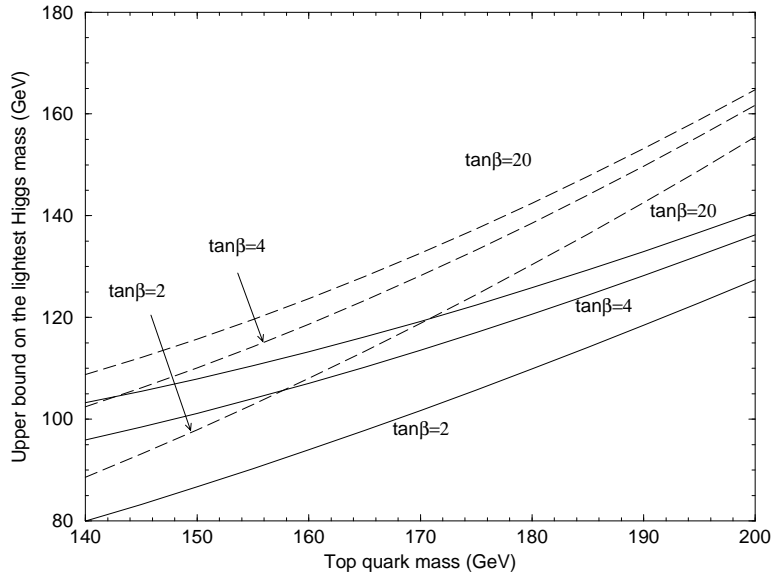,width=4in,angle=-90}}
\end{center}
\caption{The upper bound on the lightest CP-even Higgs mass in the MSSM
as a function of the top quark mass for various $\tan{\beta}$ and
two large stop mass scales. The solid (dashed) line corresponds
to $m_{stop}$=1 (10) TeV without left-right mixing of two stop states.
These masses are calculated by the method with the renormalization
group equation.\protect\cite{OYYRGE}
\label{fig:fig1}}
\end{figure}
We can see that, in the MSSM, at least one neutral Higgs-boson should
exist below 130 - 150 GeV depending on the top and stop masses.

The Higgs boson in this mass range is a target of the future
collider experiments both at LHC and $e^+ e^-$ linear colliders.
In the coming experiment at LEP II the SM Higgs boson is expected to
be discovered if its mass is below 95 GeV. Since the upper
bound exceeds the discovery limit of LEP II many efforts are made
to clarify the discovery potential of the SUSY Higgs in LHC
experiments. In this mass range the main decay mode of the SM
Higgs boson is $h\rightarrow b \bar{b}$. Unfortunately
because of QCD backgrounds we cannot use this mode
in the LHC experiments and we have to rely on 
the two photon mode whose branching ratio is $O(10^{-3})$. 
In the SUSY case its branching ratio can be even smaller,
and the search may be more difficult.
Recent study shows that it is probably possible
to get at least one signal of the SUSY Higgs sector in almost
all parameter space but we may have to wait for several
years before we find the signal.\cite{LHC} On the other hand 
an $e^+ e^-$ linear collider with $\sqrt{s}\sim300 - 500$ GeV
is a suitable place to study the Higgs boson in this 
mass region. Here we can not only discover the Higgs boson
easily but also measure various quantities, {\it i.e.}
production cross sections and branching ratios related
to the Higgs boson.\cite{Janot,Kawagoe,JLC,Hildreth}
These measurements are very important to clarify nature 
of the discovered Higgs boson and distinguish the
SM Higgs boson from Higgs particles associated with
some extensions of the SM like the MSSM.     

Other Higgs states, namely the $H, A, H^\pm$, are also important
to clarify the structure of the model. Their existence alone is proof
of new physics beyond the SM, but we may be able to distinguish the
MSSM from a general two-Higgs model through the investigation of
their masses and couplings.  In the MSSM the Higgs sector is
described by four independent parameters for which we take the
mass of the CP-odd Higgs boson ($m_A$), $\tan\beta,$ the top-quark mass
$(m_t$) and the stop mass ($m_{stop}$).
The top and stop masses enter through radiative corrections to
the Higgs potential.  Speaking precisely, there are left- and
right-handed stop states which can mix to form
two mass eigenstates; therefore more than just one parameter
is required to specify the stop sector.
In Figure 2, the masses for the $H, A$, and $ H^\pm$ are shown as a
function of $m_A$ for several choices of $\tan\beta$
and $m_{stop}$=1 TeV.
%Fig 2 (Higgs masses as a function of ma)
%
\begin{figure}
\begin{center}
\mbox{\psfig{figure=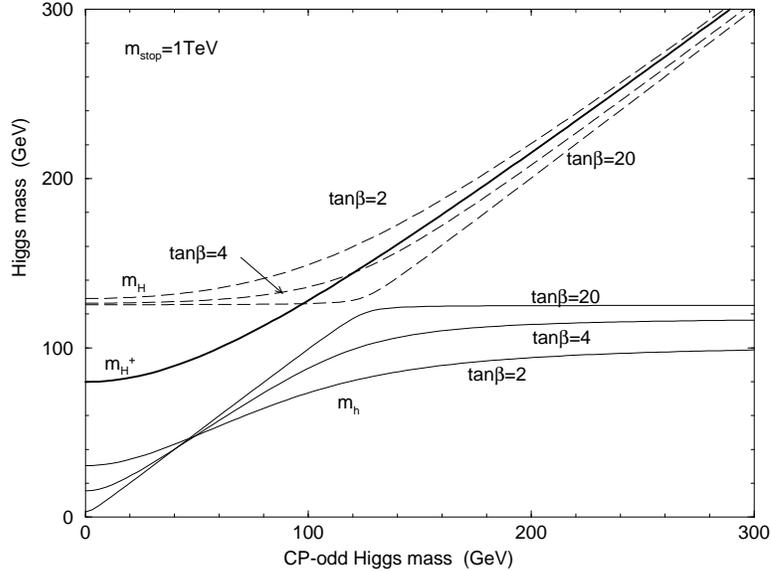,width=4in,angle=-90}}
\end{center}
\caption{The light ($h$), heavy ($H$) CP-even Higgs masses and
the charged
Higgs ($H^\pm$) mass as a function of the CP-odd Higgs ($A$) mass.
The top and stop masses are taken as $m_t$ =~170 GeV and
$m_{stop}$ =~1 TeV.
\label{fig:fig2}}
\end{figure}
We can see that, in the limit of $m_A \rightarrow \infty$, $m_h$
approaches a constant value which corresponds to the upper bound
in Figure 1. Also in this limit the
$H, A$ and $H^\pm$ become degenerate in mass.

The neutral Higgs-boson couplings to gauge bosons and fermions
are determined by the ratio of vacuum expectation values $\tan{\beta}$
and the mixing angle $\alpha$
of the two CP-even Higgs particles defined as
\begin{eqnarray}
ReH_1^0 &=& \frac{1}{\sqrt{2}}
(\upsilon\cos\beta - h\sin\alpha + H\cos\alpha)
\nonumber \\
ReH_2^0 &=& \frac{1}{\sqrt{2}}
(\upsilon\sin\beta + h\cos\alpha + H\sin\alpha).
\end{eqnarray}
For Higgs-boson production, the Higgs-bremsstrahlung process
$e^+e^- \rightarrow Zh$ or $ZH$ and the associated production
$e^+e^- \rightarrow Ah$ or $AH$ play complimentary roles.
Namely $e^+e^- \rightarrow Zh~(ZH)$ is proportional
to $\cos(\beta-\alpha)(\sin(\beta-\alpha))$, and
$e^+e^- \rightarrow Ah~(AH)$ is proportional to
$\sin(\beta-\alpha)(\cos(\beta-\alpha))$,
so at least one of the two processes has a sizable coupling.
It is useful to distinguish the following two cases when we discuss
the properties of the Higgs particles in the MSSM, namely
(i) $m_A \lsim 150$ GeV, (ii) $m_A \gg 150$ GeV.
In case (i), the two CP-even Higgs bosons can have large mixing, and
therefore the properties of
the neutral Higgs boson can be substantially different from those
of the minimal SM Higgs.  On the other hand, in case (ii), the
lightest CP-even Higgs becomes a SM-like Higgs, and the other
four states, $H, A, H^\pm$ behave as a Higgs doublet orthogonal
to the SM-like Higgs doublet.  In this region,
$\cos(\beta-\alpha)$ approaches unity and $\sin(\beta-\alpha)$
goes to zero so that $e^+e^- \rightarrow Zh$ and
$e^+e^- \rightarrow AH$ are the dominant production processes.

Scenarios for the Higgs physics at a future $e^+e^-$ linear
collider are different for two cases.  In case (i) it is possible
to discover all Higgs states with $\sqrt{s} = 500$ GeV, and the
production cross-section of the lightest Higgs boson may be
quite different from that of the SM so that
it may be clear that the discovered Higgs is not the SM Higgs.
On the other hand, in case (ii), only the lightest Higgs may
be discovered at the earlier stage of the $e^+e^-$ experiment,
and we have to go to a higher energy machine to find the heavier
Higgs bosons. Also, since the properties
of the lightest Higgs boson may be quite similar to those of
the SM Higgs boson we need precision experiments on the production
and decay of the particle
in order to investigate possible deviations from the SM.

\section{The Higgs sector in extended versions of the SUSY SM}
Although the MSSM is the most widely studied model, there are
several extensions of the SUSY version of the SM.  If we focus
on the structure of the Higgs sector, the MSSM is special because
the Higgs self-couplings at the tree level
are completely determined by the gauge coupling constants.
It is therefore important to know how the Higgs phenomenology
is different for models other than the MSSM.

A model with a gauge-singlet Higgs boson is the simplest
extension.\cite{singlet}  This
model does not destroy the unification of the three gauge coupling
constants since the new light particles do not carry the SM quantum
numbers. Moreover, we can include a term $W_\lambda = \lambda NH_1H_2$
in the superpotential where $N$ is a gauge singlet superfield.
Since this term induces $\lambda^2|H_1H_2|^2$ in the Higgs potential,
the tree-level Higgs-boson self-coupling depends on $\lambda$
as well as the gauge coupling constants.  There is no definite
upper-bound on the lightest CP-even Higgs-boson mass in this model
unless a further assumption on the strength of the coupling $\lambda$
is made.  If we require all dimensionless
coupling constants to remain perturbative up to the GUT scale
we can calculate the upper-bound of the lightest CP-even Higgs-boson
mass.\cite{singletmass} In Figure 3, the upper bound of
the Higgs-boson mass is shown as a function of the top-quark mass.
%Fig3. (NMSSM Higgs mass)
%
\begin{figure}
\begin{center}
\mbox{\psfig{figure=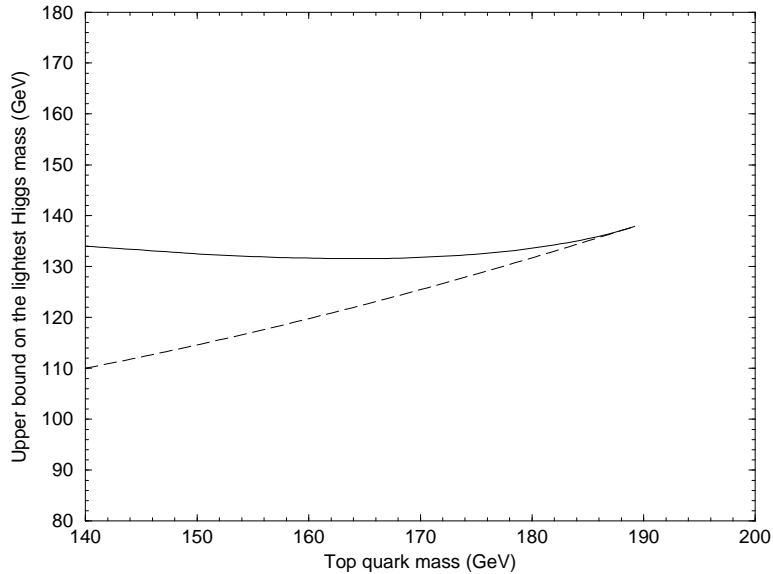,width=4in,angle=-90}}
\end{center}
\caption{The upper bound on the lightest CP-even Higgs mass
in the SUSY SM with a gauge singlet Higgs (the solid line).
The stop mass is taken as 1 TeV. The dotted line corresponds
to the upper bound in the MSSM case.
\label{fig:fig3}}
\end{figure}
In this figure we have taken the stop mass as 1 TeV and demanded
that no dimensionless coupling constant may blow up below the GUT scale
($\sim 10^{16}$ GeV).  We can see that the upper bound is given
by 130 $\sim$ 140 GeV for this choice of the stop mass.
The top-quark-mass dependence is not significant compared to
the MSSM case because the
maximally allowed value of $\lambda$ is larger (smaller) for a
smaller (larger) top mass.

 From this figure we can see that the lightest Higgs boson is at least
kinematically accessible at an $e^+e^-$ linear
collider with $\sqrt{s}\sim 300 - 500$ GeV.  This does not, however,
mean that the lightest Higgs boson is detectable.  In this model
the lightest Higgs boson is composed of one gauge singlet and two
doublets, and if it is singlet-dominated its couplings to the
gauge bosons are significantly reduced, hence its
production cross-section is too small. In such a case the heavier
neutral Higgs bosons may be detectable since these bosons have a
large enough coupling to gauge bosons. In fact we can put an
upper-bound on the mass of the heavier Higgs boson when the
lightest one becomes singlet-dominated.
By quantitative study of the masses and the production
cross-section of the Higgs bosons in this model, we can show that
at least one of the three CP-even Higgs bosons has a
large enough production cross-section in the
$e^+e^- \rightarrow Zh^o_i$ $(i =1, 2, 3)$ process to be detected
at an $e^+e^-$ linear collider with
$\sqrt{s}\sim300 - 500$ GeV.\cite{KOT}  
For this purpose we define the minimal production
cross-section, $\sigma_{min}$, as a function of $\sqrt{s}$ such that
at least one of these three $h^0_i$  has a larger production
cross section than $\sigma_{min}$ irrespective of the
parameters in the Higgs mass matrix.
The $\sigma_{min}$ turns
to be  given by one third of the SM production cross-section
with the Higgs boson mass equal to the upper-bound value.
In Figure 4 we show  $\sigma_{min}$ as a function of $\sqrt{s}$
for $m_{stop} = 1$ TeV .
%Fig4. (sigma min for NMSSM)
\begin{figure}
\begin{center}
\mbox{\psfig{figure=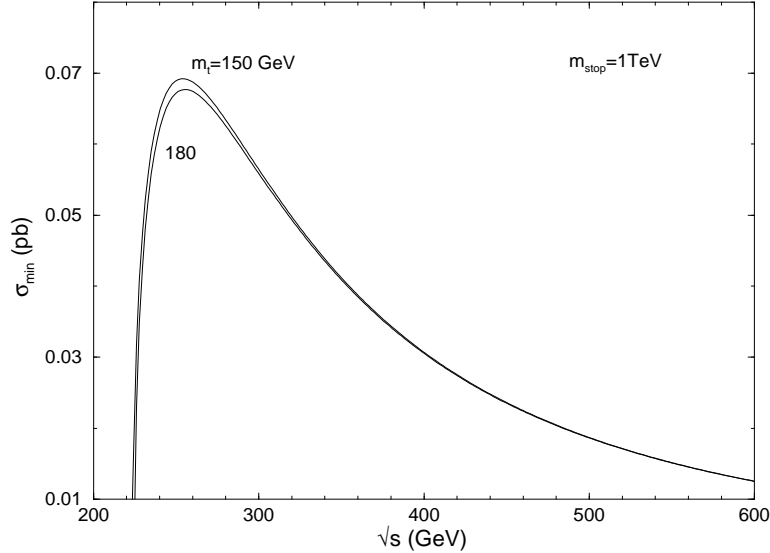,width=4in,angle=-90}}
\end{center}
\caption{Minimal production cross section, $\sigma_{min}$, for
the SUSY SM with a gauge singlet Higgs for the top mass
$m_t$=150 and 180 GeV and $m_{stop}$=1 TeV.
\label{fig:fig4}}
\end{figure}
 From this figure we can conclude that the discovery of 
at least one neutral Higgs boson is guaranteed at an $e^+e^-$
linear collider with an integrated luminosity of 10 fb$^{-1}$.

\section{Determination of the heavy Higgs mass scale
from branching measurements in the MSSM}
In the previous section we have discussed the detectability
of the Higgs boson in the SUSY SM with a gauge singlet
Higgs field. Since the situation is better for the case
of the MSSM we can show that at least one CP-even Higgs boson
of the MSSM can be discovered at the first stage of an $e^+e^-$ 
linear collider experiment where the CM energy is 
$\sim 300 -  500$ GeV.

If a Higgs boson is discovered, the next question is
to determine whether this boson is the SM Higgs boson
or a Higgs boson associated with some extension of the SM.
For this purpose it is important to know to what extent
the non-minimality of the Higgs boson can be detected
through the investigation of the production
cross-section and decay branching 
ratios.\cite{Janot,Kawagoe,JLC,Hildreth}
Here we would like to consider this problem in the context of
the MSSM, that is, we would like to know  whether
the parameters in the Higgs sector are determined by various
observable quantities related to the Higgs boson. 
Although it is possible to discover all five Higgs
states at the first stage of the linear collider experiment,
we may at first be able to find only one CP-even Higgs boson.
In such a case it is important to determine the heavy Higgs 
mass scale because the heavy Higgs bosons become targets 
of the second stage of the $e^+e^-$ linear-collider 
experiments after the beam energy is increased.

In the following analysis let us assume that only
one CP-even Higgs boson is discovered 
at the $e^+e^-$ linear-collider experiment.
The free parameters required to specify the Higgs sector 
in the MSSM can be taken
to be the CP-odd Higgs-boson mass ($m_A$), the ratio of
two vacuum expectation values ($\tan\beta$) and
masses of the top quark and the stop squark. The latter
two parameters ($m_t, m_{stop}$) are necessary to evaluate
the Higgs potential at the one-loop level.
Suppose that the lightest CP-even Higgs boson is discovered
such that its mass ($m_h$) is precisely known.
Then we can solve for one of the free
parameters, for example, $\tan\beta$, in terms of the
other parameters. Assuming the top-quark mass is well determined
by the time when the $e^+e^-$ linear
collider is under operation, the unknown parameters for
the Higgs sector are then $m_A$ and $ m_{stop}$.  The question is,
to what extent these parameters are
constrained from observable quantities such as the production
cross-section and the various branching ratios.

We show that one particular ratio of two
branching ratios,
\begin{equation}
R_{br}\equiv \frac{Br(h \rightarrow c\bar{c}) + Br(h \rightarrow gg)}
{Br(h \rightarrow b\bar{b})},
\end{equation}
is especially useful to constrain the heavy Higgs mass 
scale.\cite{Kamoshita}
In the MSSM, each of the two Higgs doublets couples to either
up-type or down-type quarks.  Therefore, the ratio of the Higgs
couplings to up-type quarks and to down-type quarks is
sensitive to the parameters of the Higgs sector, {\em i.e.} the angles
$\alpha$ and $\beta$ in Section 2.
Since the gluonic width of the Higgs boson is generated by a
one-loop diagram with an internal top-quark,
the Higgs-gluon-gluon coupling is essentially
proportional to the Higgs-top coupling.
Then $R_{br}$ is proportional to square of the ratio of
the up-type and down-type Yukawa coupling constants.  Since
the up-type (down-type) Yukawa coupling constant contains a
factor $\frac{\cos\alpha}{\sin\beta}$,
$(-\frac{\sin\alpha}{\cos\beta})$ compared to the SM coupling constant,
$R_{br}$ is proportional to $(\tan{\alpha} \tan{\beta})^{-2}$.
In Figure 5 $R_{br}$ is shown as a function of $m_A$ for 
$m_{susy}(\equiv m_{stop})=1,5$ TeV.
%Fig.5 (R_br)
%
%
\begin{figure}
\begin{center}
\mbox{\psfig{figure=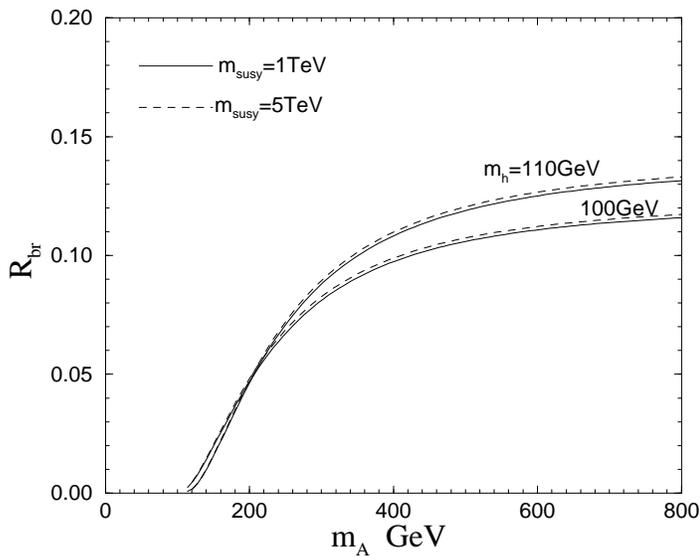,width=4in,angle=0}}
\end{center}
\caption{
$R_{br}\equiv
\protect\frac{(Br(h\to c\bar{c})+Br(h\to gg))}{Br(h\to b\bar{b})}$
as a function of $m_A$ for $m_{susy} =1, 5$ TeV and
$m_h=100, 110$ GeV.\protect\cite{Kamoshita}
The following parameters are used for the calculation of
the branching ratios: $m_t= 170$ GeV, $\bar m_c(m_c)=1.2$ GeV,
$\bar m_b(m_b)=4.2$ GeV, $\alpha_s(m_Z)=0.12$.
\label{fig:fig5}}
\end{figure}
{}From this figure we can see that $R_{br}$ is almost independent of
$m_{stop}$.  In fact, it can be shown that $R_{br}$ in the MSSM,
normalized by $R_{br}$ in the SM, is
approximately given by,
\begin{equation}
\frac{R_{br}(MSSM)}{R_{br}(SM)}\approx \left(\frac{m_h^2 - m_A^2}
{m_Z^2 + m_A^2}\right)^2
\end{equation}
for $m_A \gg m_h \sim m_Z$.  
Measuring this quantity to a good accuracy is therefore 
important for constraining the scale of the heavy
Higgs mass. Note that $R_{br}$ approaches the SM value in the
large $m_A$ limit. We can see that $R_{br}$ is reduced by 20$\%$
even for $m_A = 400$ GeV. By simulation study for 
$e^+e^-$ linear collider experiments it is shown that  
the sum of the charm and gluonic branching ratios 
can be determined reasonably well.
The statistical error in the determination of $R_{br}$ after two
years at an $e^+e^-$ linear collider with $\sqrt{s} = 300$ GeV
is 17$\%$.\cite{Nakamura}
We also need to know the theoretical ambiguity of the
calculation of the branching
ratios in $h \rightarrow b\bar{b}, c\bar{c}, gg$.
At the moment the theoretical error in the calculation of $R_{br}$
is estimated to be rather large ($\sim$ 20$\%$) due to
uncertainties in $\alpha_s$ and $m_c$.\cite{Kamoshita,DSZ}
But these uncertainties can be reduced in future from both
theoretical and experimental improvements. 

\section{Conclusions}
I have reviewed some aspects of the Higgs physics 
in the SUSY SM. I have shown that an future $e^+ e^-$ linear 
collider is an ideal place to study the SUSY Higgs sector.
At earlier stage of the experiment with
$\sqrt{s}\sim$ 300 - 500 GeV, it is easy to find a light
Higgs boson predicted in SUSY standard models.
In particular, both in the MSSM and the SUSY SM with
a gauge singlet Higgs, at least one of neutral Higgs bosons
is detectable. More importantly,
detailed study on properties of the Higgs boson is possible
at an $e^+ e^-$ linear collider through measurements of various 
production cross-sections and branching ratios. As an example 
we show that the measurement of Higgs couplings 
to $c\bar{c}$/$gg$/$b\bar{b}$ gives us important information 
on the Higgs sector of the MSSM. 
It is therefore very important to build an $e^+ e^-$ linear 
collider along with LHC, then combining both results 
we will be able to clarify the Higgs sector
of the SM and explore physics beyond the SM such as the SUSY SM.

%\section{Acknowledgements}

%\section{References}
%References in the bibliography should be referred to in the text
%by a superscript number without parentheses or brackets. All
%references should be organized to provide initials and last name
%of the author(s), title of publication (in italics), volume (in
%boldface), year of publication of paper in the journal/book and
%page numbers, e.g.,
%
%\begin{thebibliography}{9}
%\bibitem{Don/Ho} J. F. Donoghue and B. R. Holstein, {\it Phys.
%Rev.} {\bf D25} (1982) 2015.
%\bibitem{Coh/An} M. L. Cohen and P. W. Anderson, in {\it
%Superconductivity in d- and f-Band Metals}, ed. D. H. Douglas
%(AIP, New York, 1972).
%\bibitem{Kre} H. Krebs, {\it Fundamentals of Inorganic Crystal
%Chemistry} (McGraw-Hill, London, 1968), p. 160.
%\end{thebibliography}

\end{document}